\documentclass[aps,pre,twocolumn,groupedaddress,showpacs]{revtex4}
\usepackage{epsf}
\newcommand{\lab}[1]{\label{#1}}
\newcommand{\re}[1]{(\ref{#1})}
\newcommand{\ci}[1]{\cite{#1}}
\newcommand{\be}{\begin{equation}}
\newcommand{\ee}{\end{equation}}
\newcommand{\ba}{\begin{eqnarray}}
\newcommand{\ea}{\end{eqnarray}}
\date{}

\begin{document}
\date{}
 \title{QUANTUM CHAOS IN THE HEAVY QUARKONIA}
\author{D.U.Matrasulov$^{(a)}$,
F.C.Khanna$^{(a)}$, U.R.Salomov$^{(b)}$
\thanks{e-mail: lps@mnet.uz}}
\date{\today}
\affiliation{
(a) Physics Department University of Alberta\\
 Edmonton Alberta, T6G 2J1 Canada\\
 and TRIUMF, 4004 Wersbrook Mall,\\ Vancouver, British Columbia, Canada, V6T2A3\\
(b) Heat Physics Department of the Uzbek Academy of
Sciences,\\
28 Katartal St.,700135 Tashkent, Uzbekistan}

\begin{abstract}
The binding energy spectra of the heavy quarkonia are
 calculated by solving the Schr\"odinger equations with Coulomb plus confining potentials.
Statistical properties of the obtained spectra are studied by plotting nearest level spacing distribution histograms.
\end{abstract}
\pacs{ 05.45 Mt; 14.40.Lb; 14.40.Nd; 12.39.Pn}
\maketitle
\section{Introduction}
Level statistics, i.e. the  statistical theory describing the fluctuation properties of the quantum spectra was originally developed in
nuclear physics in early fifties.
Pioneering work dates back to Wigner \ci{Wig51} who gave the first detailed theoretical analysis of the nuclear spectra
from the viewpoint of random matrix theory.
Further developments of this subject were carried out  by Porter \ci{Porter}, Mehta \ci{Mehta}  and Dyson \ci{Dyson}.
In particular, it has been established that the spectra of the complex quantum many-body systems have a
 behaviour similar to that of the random matrices \ci{Brod}.

Later  the development of the statistical spectroscopy was stimulated by quantum chaology which
arose  in early eighthies as a quantum theory of classical chaotic systems.
It was established that  quantized spectra of classical chaotic system spectra have the same statistical properties as those  of
the random matrices.
It is well known, that the classical chaotic dynamics is characterised by an exponential divergence of the neighboring trajectories in
phase-space.
How far the
complexity of the classical motion is reflected in the wave dynamics of the corresponding
quantum system (described by a linear wave equation) is an interesting aspect of the study of deterministic non-linear dynamics.
 Numerical evidence has shown that
the statistical properties of the energy spectra of the quantum systems depend crucially on
the degree of chaos  in the underlying classical dynamics.
Namely, the spectrum of the quantum operator, describing classically non-integrable
systems, has the same statistical properties as those of random matrices.
It has been established that certain distribution
functions and fluctuation measures for a wide variety of complex nuclear spectra follow a universal
behavior and that  moreover  they agree very well with the results obtained by ensemble averaging
over certain matrix ensembles with randomly distributed matrix elements. In particular, for systems with
time-reversal symmetry,  good agreement was found with a Gaussian orthogonal ensemble(GOE)  \ci{Eckh88, Gutz90}.
It has also been established, that for integrable systems the nearest neighbor spacing distribution
is dominated by level clustering and is expected to be Poisson-like. Later, this has been
confirmed in a number of studies of model systems like billiard balls, simple rotators etc.\ci{Eckh88, Bohig, Selig}
as well as in realistic systems such as hydrogen atom in a magnetic field \ci{Wint, Del}.
Details of the progress  of level statistics and random matrix theory are given in a recent review \ci{Guhr}.
Here  we deal with the spectral statistics of the quark-anitquark systems(quarkonia).
Quarkonium spectrosopy has been the subject of extensive experimental as well as theoretical studies  \ci{AL}-\ci{Mukh93}.
Theoretical description of quarkonium properties can be done quantum mechanically as well as within the QCD approach.
Quantum mechanical description of hadrons is reduced to solving  wave equations with Coulomb plus a confining potential.
Considerable progress has been made in the calculation of the spectra of quarkonium \ci{Quig}-\ci{Zhu} and baryons \ci{Mukh93}.
At present a large amount of theoretical and experimental data is available.
 Recently energy fluctuations and quantum chaos in hadrons and QCD has become a subject of theoretical studies \ci{Hal}-\ci{Gu}.
In partuclar, it has been found that QCD is governed by quantum chaos in both confinement and deconfinement
phases \ci{Hal,Bit}.
The statistical analysis of the measured meson and baryon spectra shows that there is quantum chaos phenomenon
in these systems \ci{Pasc}.
The  study of the charmonium spectral statistics  and its dependence on color screening has established quantum chaotic behaviour
of this system \ci{Gu}. It was claimed that such a behaviour could be the reason for  $J/\Psi$ supression \ci{Gu}.

In the present work we concentrate our attention on level fluctuations and quantum chaos in heavy quarkonia due to the confining forces.
It is well known that many realistic and model confined dynamical systems such as an atom in a magnetic field,
particle motion in resonators, quantum dots,
billiards   exhibit chaotic dynamics.  Quarks in the hadrons are also confined systems, and to some
extent are equivalent to  motion in billiards(domain with hard walls where elastic scattering of the particle will occur )
\ci{Bun, Ber, Zas}. Depending on the shape of the billiard domain the  motion can be regular or chaotic \ci{Ber}.
If one uses for quark- antiquark potential Coulomb plus harmonic oscillator potential, the  system  becomes equivalent to the atom  in a magnetic
filed, which is also  described by a Coulomb plus harmonic oscillator potential whose dynamics is chaotic in classical and quantum cases \ci{Del,Wint}.
An analysis of the quarkonium would suggest that the dynamics of such a system
could be chaotic in classical as well as in quantum cases . Therefore the present paper may be viewed as an attempt to use
well-known technique to the quarkonium systems to establish their chaotic behaviour.

Charmonium and bottomonium spectral statistics based on the solution of the Schr\"odinger equation for this system is considered .
Two types of potentials for quark-antiquark interaction are considered: Coulomb plus linear and Coulomb plus harmonic oscillator potentials.
Solving the Schr\"odinger equation with these potentials by the scaling variational method we obtain the energy levels of the quarkonium system.
This paper is organized as follows: In the next section we solve the Schr\"odinger equation analytically using the scaling variational method and find the mass sectra of charmonium and bottomonium.
In section 3 we calculate the statistical properties of the spectra of charmonium and bottomonium and in section 4 some conclusions and final remarks are presented.

\section{Schr\"odinger equation for quarkonium}
To obatain the quantum spectra the following Schr\"odinger equation is solved
\be
(-\frac{1}{2\bar M}\triangle-\frac{Z}{r}+\lambda r^k)\Psi=E\Psi,
\lab{eq1}
\ee

where
 $Z =\frac{4}{3}\alpha_s$, with $\alpha_s$ being  the strong coupling constant, $\lambda$ is the parameter of the confining potential,
$k=1,2$ and  $\bar M= m_q/2$ is the reduced mass of quarkonium (for  simplicity we consider quarks with the same flavor).
To solve this equation we use the scaling variational method. This method  \ci{Castro}
was shown  to work well for Coulomb plus confining type potentials. Recently this method was used for the calculation of doubly heavy
baryon spectra \ci{Mat00} and heavy-light meson spectra \ci{Matr02}.
For analysis,  eq. \re{eq1} may be written as
\be
(H_0+H')\Psi_{nl} = E_{nl}\Psi_{nl},
\ee
where
\be
H_0= -\frac{1}{2}\triangle-\frac{Z}{r},
\lab{eq3}
\ee
and
\be
H'=\lambda r^k.
\lab{eq3}
\ee

According to the scaling variational method we choose the  trial wave function of the form
\be
\Psi_{nl}(b, r) = b^{3/2}\phi_{nl}(b r),
\ee

where $\phi_{nl}(b r)$ are the the eigenfunctions of the Hamiltonian $H_0$, i.e.
$$
H_0\phi_{nl} =\varepsilon_{nl}\phi_{nl},
$$

satisfying the the normalization condition
$$
<\phi_{nl}|\phi_{n'l'}> = \delta_{nn'} \delta_{ll'}.
$$
and the eigenvalues are given as
$$
\varepsilon_{nl} = -\frac{Z^2}{2n^2}.
$$

Then the eigenvalues of  eq.\re{eq1} can be calculated as
\be
E_{nl}  = \epsilon_{nl}(b_0),
\lab{spec}
\ee

where
$$
\epsilon_{nl}(b) =  b^{3}<\phi(b,r) |H_0+H'|\phi(b,r) >,
$$
and $b_0$ is found by minimizing this energy, i.e.
\be
\frac{\partial \epsilon}{\partial b} = 0.
\lab{ex}
\ee

The quantity $\epsilon_{nl}(b)$ may  be calculated as:

$$
\epsilon_{nl}(b)=n^{-2}(\frac{b^2}{2}-Zb-\frac{qb^{-k}}{k}),
$$
where
$$
q=k\lambda n^2<\phi_{nl}|r^k|\phi_{nl}>.
$$

These matrix elements are calculated analytically \ci{Lan}.
For $k=1$ we have
$$
q=\lambda n^2 \frac{1}{2}(3n^2-l(l+1)).
$$
where $n$ and $l$ are the principal and orbital quantum numbers, respectively.
Then the extremum condition gives

$$
b_0=\frac{Z}{3}+\frac{2^{\frac{1}{3}}Z^2}{3(27q+2Z^3+3^{\frac{3}{2}}(q(27q+4Z^3))^{\frac{1}{2}})^{\frac{1}{3}}}+
$$
$$
\frac{3(27q+2Z^3+3^{\frac{3}{2}}(q(27q+4Z^3))^{\frac{1}{2}})^{\frac{1}{3}}}{2^{\frac{1}{3}} 3}.
$$
For $k=2$ we have
$$
q= 5n^2+1-3l(l+1).
$$

The extremum condition \re{ex} leads to an equation for defining $b_0$
$$
b_0^{4}-Zb_0^{3}-q=0.
$$

Solving this equation we get   $b_0$  that leads to  the energy eigenvalues for
the case $k=2$.
Thus the solutions of the Schr\"odinger equation for Coulomb plus confining potentials are obtained analytically
 using the scaling variational method.

\section{Spectral statistics}
Now an analysis of  the  statistical  properties of the spectra is to be considered.
One of the main characteristics of the statistical properies of the spectra is  the level spacing distribution \ci{Eckh88, Gutz90, Bohig} function.
We calculate the nearest-neighbor level-spacing distribution for quarkonium spectra.
The nearest neighbor level spacings are defined as $S_i = \tilde E_{i+1}- \tilde E_i$,
where $\tilde E_i$ are the energies of the unfolded levels, which are obtained by the following way:
The spectrum $\{E_i \}$  is separated into smoothed average part and fluctuating parts. Then the  number of the levels below $E$ is counted
and the following staircase function is defined:
$$
N(E)  = N_{av}(E) +N_{fluct}(E).
$$

The unfolded spectrum is finally obtained with the mapping
$$
\tilde E_i = N_{av}(E_i).
$$

Then the nearest level spacing distribution function $P(S)$ is defined as the probability of  $S$  lying within the  infinitesimal interval $[S, S+dS]$.

For the quantum systems which are chaotic in the classical limit this  distribution function is the same as that of the random matrices \ci{Eckh88,Guhr}.
For  systems which are regular in the classical limit its behaviour is close to a Poissonian distribution function.
This distribution is usually taken to be a Gaussian with a parameter $d$:
$$
P(H) \sim exp(-Tr\{HH^+\}/2d^2),
$$

and the random matrix ensemble corresponding to this disribution is called the Gaussian ensemble.
For Hamiltonians invariant under rotational and time-reversal transformations the  corresponding ensemble
of matrices is called Gaussian orthogonal ensemble (GOE).
It was established \ci{Bohig, Selig,Ber85} that GOE describes  the statistical fluctuation properties of a quantum system whose classical analog is
completely chaotic.
The nearest neighbor level spacing  distribution for GOE is described by  Wigner distribution:
\be
 P(S) = \frac{1}{2}\pi S exp(-\frac{1}{4}\pi S^2).
 \lab{wig}
\ee
 The common way to study of the level statistics is to compare the calculated nearest-neighbor level-spacing distribution histogram with the
 Wigner distribution.

 For systems whose classical motion is neither regular nor fully chaotic (mixed dynamics)  the level spacing distribution will be
 intermediate between the Poisson and GOE limits. Several  empirical functional forms for the distribution were suggested for this case \ci{Guhr}.
If the Hamiltonian is not time-reversal invariant, irrespective of its behavior under rotations, the Hamiltonian matrices are  complex Hermitian and
the corresponding ensemble is called Gaussian unitary ensemble (GUE).
If the Hamiltonian of the systems is time-reversal invariant but not invariant under rotations, then the corresponding ensemble is called the Gaussian
symplectic ensemble.
Quarkonium system is time reversal and rotational invariant. Therefore  in the case of chaotic motion its nearest-neighbor level spacing
distribution is expected to be GOE-type.
Now we carry out a detailed analysis of the spectra to establish the nearest-neighbor level spacing distribution function to determine the existence
of quantum chaos in our system.

 The following values of the $c$ and $b$ quark masses and potential parameters
are choosen in these claculations:\\
For charmonium:\\  $m_c = 1.486 GeV\;$,
$\alpha_s = 0.32$, $\lambda =0.2GeV^2, 0.4GeV^2, 0.6GeV^2 $.\\
For bottomonium :\\
 $m_b = 4.88 GeV\;$,
$\alpha_s = 0.22$, $\lambda =0.4GeV^2, 0.6GeV^2, 0.8GeV^2$.
First 1000 eigenavleus are used to plott each histogram. In our claculations we vary only the principal quantum number $n$, keeping
the orbital quantum number $l$   constant.

Figs. 1  and 2 show the  level spacing distribution for charmonium
and bottomium, respectively  for Coulomb plus linear potential ($k=1$) case, and in Fig.s 3 and 4 the level spacing distributions are  plotted for
Coulomb plus harmonic oscillator potential ($k=2$) case,
at various values of the confining potential parameter.
It is clear that  for the smaller
values of the parameter $\lambda$ this distribution is close to GOE,  that means  the quantum system is becoming  chaotic.
It allows us to estimate the values of  $\lambda$  where a  transition from regular to chaotic behaviour will occur.
Furthermore we find  that for the same values of  $\lambda$ the bottomonium motion is more regular than that of the charmonium.
Comparison of the level spacing distribution  for $c\bar c$ and $b \bar b$  systems shows us that the transition from regular
 to chaotic motion for these systems occurs at different values of $\lambda$.
 So, for $c\bar c$ system (Figs.1 and 2) we have GOE-type distributions at $\lambda=0.2GeV^2$ and $\lambda =0.4 GeV^2$ and Poissonian
  distribution at $\lambda =0.6 GeV^2$, while the motion of $b\bar b$ system is still chaotic at $\lambda=0.6GeV^2$ (Figs.3 and 4).
   At  $\lambda=0.4GeV^2$ and $\lambda =0.8GeV^2$ the level spacing distribution for bottomonium
   is GOE and Poisson type, respectively. Thus it may be concluded from this comparison that the motion of $b\bar b$ system
   is more chaotic than that of the $c\bar c$ system.

Indeed, if one writes the Schr\"odinger equations for charmonium and bottomonium in the units $m_q=\hbar =1$, the value of the
confining potential  for bottomonium is much  smaller than that of charmonium.
The chaotic nature of the motion in the quarkonium potential can be also understood from the following fact:
formally the Hamiltonian for  Coulomb plus harmonic oscillator potential is equivalent to the Hamiltonian of an atom in an uniform magnetic field,
which is a nonintegrable system. It is well known,  that the Hamiltonian of an atom in an uniform magentic field can be reduced to
the  Hamiltonian of coupled nonlinaer oscillators \ci{Del}, which exhibits chaotic dynamics.
With this analogy the change in the confining potential parameters to some extent is equivalent to the change of the shape
of billiard domain.

\section{Conclusions}
Summarizing, in this work we have applied the technique well-known in nuclear physics, 
atomic physics and other areas of physics to treat quantum chaos in heavy quarkonia, using for their description the
Schr\"odinger equation with Coulomb plus confining potentials, a non-integrable system.
It should be emphasized that this system is quite similar to the case of an atom in a constant magnetic field.
It is well known that such a system is non-integralble and exhibits chaotic dynamics in both classical as well
as quntum cases\ci{Del,Wint}.
The analysis of the nearest level spacing distributions of these systems
shows, that
for some smaller values of the confining potential parameter the motion of the heavy quarkonium is chaotic while for larger values  of
$\lambda$ the level spacing distribution is a Poisson type, that implies that the  motion is becoming regular.
Previously  \ci{Gu} the  quantum chaos is
related to the color screening parameter, while the present  work treats
the transition from regular to a chaotic motion   related to a confining potential parameters.
The quarkonia can exhibit regular as well as chaotic dynamics depending on  the value of the confining force.
Study of the level statistics and quantum chaos in hadrons could play the same role in particle physics as
it did   in the case of nuclear physics.
Therefore quantum chaology of hadrons could become one of the exciting topics in  particle physics.
More comprehensive  analysis of the statistical properties of the heavy quarkonia spectra performed by solving the Dirac
equation with Coulomb plus a confining potential is needed. This would establish the role played by relativistic effects in the approach
to a level-spacing distribution.

\section*{Acknowledgements}
The work of DUM is supported  by NATO Science Fellowship of Natural
Science and Engineering Research Council of Canada (NSERCC).
The work of FCK is supported by NSERCC.
The work of URS is spupported by a Grant of the  Uzbek Academy of Sciences  (contract No 33-02).

\end{document}